\documentclass[%
 amsmath,amssymb,
 aps,
]{revtex4-1}

\usepackage{graphicx}
\usepackage{dcolumn}
\usepackage{bm}

\begin{document}

\title{Hawking Radiation Is Pure}
\author{Jingbo Wang}
\email{Essay written for the Gravity Research Foundation 2020 Awards for Essays on Gravitation.}
\affiliation{shuijing@mail.bnu.edu.cn\\
Institute for Gravitation and Astrophysics, College of Physics and Electronic Engineering, Xinyang Normal University, Xinyang, 464000, P. R. China}
 \date{\today}
\begin{abstract}
Information loss paradox is still a challenging problem in theoretical physics. In this essay, for statics BTZ black holes and Schwarzschild black holes, we propose a simple solution based on the similarity between black holes and topological insulators. That is, the Hawking radiation is pure due to the entanglement between the left-moving sector and right-moving sector of the Hawking radiation. And this entanglement may be detected in an analogue black hole in the near future.
\end{abstract}
\maketitle
\section{Introduction}
One important contribution of Stephen Hawking is the discovery of the Hawking radiation \cite{hawk1,hawk2}, that is, the black hole can radiate with a black body temperature $T_H$. The thermal nature of those radiation lies at the heart of the ``information loss paradox" \cite{info1}. It says that the evaporation of a black hole will break the unitary, or in other words, a pure state can evolve into a mixed state. Up to now, there are many proposals for its resolution. For some recent reviews, see Ref.\cite{info2,info3}.

In this essay we propose a simple solution to this paradox: the Hawking radiation is pure even through it appears as thermal. This solution is based on our early claim: the black hole can be considered as a kind of the topological insulator. For the BTZ black hole in three dimensional spacetime this claim is tested in Ref.\cite{wangbms2,wangplb1}. The boundary modes on the horizon of a BTZ black hole can be described by two chiral massless scalar fields with opposite chirality \cite{whcft1}. This is the same as a topological insulator in three dimensional spacetime (also called quantum spin Hall state) \cite{tibf1}. The topological insulator has the $W_{1+\infty}$ symmetry group which contains the near horizon symmetry group of the BTZ black hole. The microstates of the BTZ black hole are identified with the quantum states of those chiral scalar fields. For higher dimensional black holes, such as Kerr black holes in four dimension, the boundary modes can be described by boundary BF theory \cite{wmz}, which is also the same as higher dimensional topological insulators. From the boundary BF theory, one can construct a scalar field theory. Just like the BTZ black hole, we can also identify the quantum states of the scalar field with the microstates of Kerr black holes \cite{wangbms4}.

\section{Examples}
It is a little counterintuitive that how a pure state appears as thermal. Actually we can understand this if the entanglement is took into account. Let us take some examples \cite{harl1}.

The first example is the vacuum state in Minkowski spacetime with Rindler decomposition \cite{uw1}. The vacuum state can be written as
\begin{equation}\label{1}
  |0_M>=\frac{1}{\sqrt{Z}}\sum_i e^{-\pi \omega_i} |i^*>_L \otimes |i>_R,
\end{equation}
where L/R represent the left and right Rindle wedge. $|i>_R$ are eigenstates of the right Rindle Hamiltonian $H_R$ and $|i^*>_L=\Theta^+ |i>_R$ with $\Theta$ the CPT operator. The reduced density matrix for the right wedge is
\begin{equation}\label{2}
  \rho_R=\frac{1}{Z}\sum_i e^{-2\pi \omega_i} |i>_R <i|_R,
\end{equation}
which is just a thermal state. So a pure state can appear as thermal if only part of it is considered.

The second example is closely related to the first one, that is the Hartle-Hawking state for an eternal Schwarzschild black hole \cite{hh1,hh2}. The state is given by
\begin{equation}\label{3}
  |\Psi_{HH}>=\frac{1}{\sqrt{Z}}\sum_i e^{-\beta E_i/2} |i^*>_L \otimes |i>_R.
\end{equation}
Here $i$ labels eigenstates of the Schwarzschild Hamiltonian in the left and right exteriors. The reduced density matrix for the right side is
\begin{equation}\label{4}
   \rho_R=\frac{1}{Z}\sum_i e^{-\beta E_i} |i>_R <i|_R,
\end{equation}
which is a thermal state at Hawking temperature. One can construct a mixed state $\rho=\rho_L \otimes \rho_R$ which is indistinguishable from the Hartle-Hawking state for local observer who stays at left or right exteriors. It is the entanglement that make the Hartle-Hawking state pure, but this entanglement is undetectable since the left and right exteriors of the eternal Schwarzschild black hole are causal disconnected.

The third example is the ground state for topological quantum states which have a chiral edge state, such as a general quantum Hall state \cite{lh1,qi1}. Separating the whole region into two parts $A$ and $B$ with a common boundary $C$. When the coupling between those two parts is ignoring, there are chiral and anti-chiral edge states propagating along the boundary. Due to the entanglement between $A$ and $B$, the reduced density matrix of the ground state for right-moving sector is thermal. The ground state $|G>$ for an abelian fractional quantum Hall state with fixed topological sector is given by \cite{qi1}
\begin{equation}\label{5}
  |G>=\frac{1}{\sqrt{Z}}\sum_n e^{-\tau_0 (H_R+H_L)} |k(n)>_R \otimes |-k(n)>_L.
\end{equation}
Here $\tau_0$ is the extrapolation length which is determined by the energy gap. $H_L$ and $H_R$ denote the Hamiltonian for the left-moving and right-moving edge sectors. $k(n)$ denotes the momentum of the state. Notice that due to the chirality the right-(left-)moving edge sector only contains excitations with positive (negative) momentum. The reduced density matrix for the right-moving sector is
\begin{equation}\label{6}
  \rho_R=\frac{1}{Z}\sum_n e^{-4 \tau_0 H_R} |k(n)>_R <k(n)|_R,
\end{equation}
which is a thermal state. This example will play important role in our following analysis for the black hole.
\section{Static BTZ black hole and Schwarzschild black hole}
In this section we consider black holes. Firstly we consider the BTZ black hole in three dimensional spacetime. The horizon separate the whole spacetime into two regions: the exterior and interior of the black hole. Similar to the quantum Hall states, the boundary modes on the horizon contain left-moving sector and right-moving sector. The Hamiltonian and angular momentum for the left-moving and right-moving sectors are as follows \cite{wangbms4},
\begin{equation}\label{10}\begin{split}
  \hat{H}_R=\sum_{k>0} \frac{k}{r_+}\hat{a}_k^+ \hat{a}_k,\quad \hat{H}_L=\sum_{k<0} -\frac{k}{r_+}\hat{a}_k^+ \hat{a}_k,\quad k \in Z\\
  \hat{J}_R=\sum_{k>0} k \hat{a}_k^+ \hat{a}_k,\quad \hat{J}_L=\sum_{k<0} k \hat{a}_k^+ \hat{a}_k.
\end{split}\end{equation}
The entropy for the right/left-moving sector is
\begin{equation}\label{8}
  S_{R/L}=\frac{\pi}{4 G}(r_+\pm r_-).
\end{equation}
We consider those entropy as entanglement entropy of Hawking radiation for right/left-moving sector. To make the whole Hawking radiation a pure state, a necessary condition is that the entanglement entropy for two sectors should be equal, that is, $r_-=0$. So in the following we will consider the static BTZ black holes.

In Ref.\cite{wanghk1} the Hawking radiation from the boundary scalar field was investigated. We found that the number distribution of the Hawking radiation for rotating BTZ black holes is the mixture of thermal radiation of right/left-moving sector,
\begin{equation}\label{9}\begin{split}
  <N(\omega_k)>=\frac{1}{e^{\beta_H (\omega_k-k \Omega_H)}-1}=\frac{1}{e^{\beta_R \omega_k}-1}H(k)+\frac{1}{e^{\beta_L \omega_k}-1}H(-k)=<N^R(\omega_k)>+ <N^L(\omega_k)>,
\end{split}\end{equation}
where $H(k)$ is the Heaviside step function, $\Omega_H=\frac{r_-}{r_+ L}$ the horizon angular velocity, and $\beta_R,\beta_L$ are inverse temperatures for right-moving and left-moving sectors. For static BTZ black holes, one has $\Omega_H=0$ and $\beta_R=\beta_L=\beta_H$.

The thermal radiation for right-moving sector $<N^R(\omega_k)>$ corresponds to a thermal density matrix,
\begin{equation}\label{11}
  \rho_R=\frac{1}{Z}\sum_{k>0} e^{-\beta_R H_R} |k,k>_R <k,k|_R,
\end{equation}
where the first $k$ represent the energy and the second angular momentum of the state.

The entanglement between the exterior and the interior of the black hole leads to the entanglement between the left-moving edge state and right-moving edge state. So similar to the fractional quantum Hall state case (\ref{5}), the state for whole Hawking radiation is a pure state
\begin{equation}\label{12}
  |G>=\frac{1}{\sqrt{Z}}\sum_{k>0} e^{-\beta_H (H_R+H_L)/2} |k,k>_R \otimes |k,-k>_L,
\end{equation}
even through it appears as thermal. Actually this state is an example of maximally entangled state, and for any right-moving state with angular momentum $k$ there exists a left-moving edge state with opposite angular momentum $-k$. It is this entanglement that make the state pure. And this entanglement can be detected in principle, unlike the case for Hartle-Hawking state for the eternal Schwarzschild black hole. 

The same procedure can be applied to the Schwarzschild black hole, since it has the similar structure as the statics BTZ black hole.
\section{Conclusion}
In this essay we propose a simple solution to the information loss paradox for static BTZ black holes and Schwarzschild black holes, based on the similarity between black holes and topological insulators. The entanglement between the left-moving edge state (with angular momentum $k<0$) and right-moving edge state (with angular momentum $k>0$) is the key to keep the finial state a pure state. Since the Hawking radiation is pure for all the time, there is no information loss.

For Kerr black holes, the above procedure need some modifications. A possible approach is that firstly a Kerr black hole evolves into a Schwarzschild black hole by spontaneous quantum superradiance emission effect. Then this Schwarzschild black hole radiates the Hawking radiation which is pure all the time. But the detail need further investigation.

It is impossible to detect the Hawking radiation for real black holes due to technical limitations. But it was reported that the analogue Hawking radiation was observed in an analogue black hole \cite{abh2}. We hope that the entanglement between the left-moving and right-moving edge states (\ref{12}) can also be observed in this type experiment in the near future.

\acknowledgments
 This work is supported by Nanhu Scholars Program for Young Scholars of XYNU.


\end{document}